\documentclass[num-refs, 12pt]{wiley-article}
\usepackage{CJK}
\usepackage{graphicx}
\usepackage{placeins}
\usepackage[space]{grffile}
\usepackage{latexsym}
\usepackage{textcomp}
\usepackage{longtable}
\usepackage{tabulary}
\usepackage{booktabs,array,multirow}
\usepackage{amsfonts,amsmath,amssymb}
\usepackage{natbib}
\usepackage{url}
\usepackage{hyperref}
\hypersetup{colorlinks=false,pdfborder={0 0 0}}
\usepackage{etoolbox}
\usepackage{multirow}
\usepackage{float}

\makeatletter

\patchcmd\@combinedblfloats{\box\@outputbox}{\unvbox\@outputbox}{}
\makeatother
\newif\iflatexml\latexmlfalse

\AtBeginDocument{\DeclareGraphicsExtensions{.pdf,.PDF,.eps,.EPS,.png,.PNG,.tif,.TIF,.jpg,.JPG,.jpeg,.JPEG}}

\usepackage[utf8]{inputenc}
\usepackage[english]{babel}

\usepackage{siunitx}

\iflatexml

\else

\paperfield{}
\abbrevs{HD, Huntington's disease; MLE, Maximum Likelihood Estimation; DCL, Diagnose Confidence Level.}
\corraddress{Ying Zhang}
\corremail{ying.zhang@unmc.edu}

\fundinginfo{DHHS/NIH/NIGMS, Grant/Award Number: U54GM115458}
\fi

\papertype{Original Article}

\title{Joint analysis for multivariate longitudinal and event time data with a change point anchored at interval-censored event time}

\author[1]{Yue Zhan}
\author[1]{Cheng Zheng}
\author[1]{Ying Zhang}

\affil[1]{Department of Biostatistics, College of Public Health, University of Nebraska Medical Center, Omaha, Nebraska, United States}

\runningauthor{Yue Zhan}

\begin{document}
\begin{CJK}{UTF8}{gbsn}
\maketitle
\selectlanguage{english}
\begin{abstract}
Huntington's disease (HD) is an autosomal dominant neurodegenerative disorder characterized by motor dysfunction, psychiatric disturbances, and cognitive decline. The onset of HD is marked by severe motor impairment, which may be predicted by prior cognitive decline and, in turn, exacerbate cognitive deficits. Clinical data, however, are often collected at discrete time points, so the timing of disease onset is subject to interval censoring. To address the challenges posed by such data, we develop a joint model for multivariate longitudinal biomarkers with a change point anchored at an interval-censored event time. The model simultaneously assesses the effects of longitudinal biomarkers on the event time and the changes in biomarker trajectories following the event. We conduct a comprehensive simulation study to demonstrate the finite-sample performance of the proposed method for causal inference. Finally, we apply the method to PREDICT-HD, a multisite observational cohort study of prodromal HD individuals, to ascertain how cognitive impairment and motor dysfunction interact during disease progression.

\textbf{Keywords} --- Joint model, Survival analysis, Longitudinal biomarker, Change point, Interval-censored data%

\end{abstract}%

\section{Introduction}
Huntington's disease (HD) is an inherited, autosomal dominant neurodegenerative disorder characterized by progressive motor dysfunction, cognitive decline, and psychiatric disturbances. The disease is caused by an abnormal expansion of CAG trinucleotide repeats in the \textit{HTT} gene on chromosome 4, which leads to the production of a mutant huntingtin protein that disrupts normal cellular function ~\cite{walker2007huntington,ross2014}. Currently, there is no effective treatment available for HD patients. 

Huntington's disease encompasses motor, cognitive, and psychiatric manifestations, but the clinical diagnosis of onset is determined primarily by the emergence of motor symptoms~\cite{paulsen2008detection}. However, cognitive impairment has been found to appear prior to HD motor diagnosis~\cite{duff2010mild,paulsen2011cognitive,paulsen2013cognitive,stout2012evaluation,zhang2021mild}. Zhang et al.~\cite{zhang2021mild} have thoroughly studied mild cognitive impairment (MCI) in various cognitive domains characterized by Harrington et al.~\cite{harrington2012cognitive}, which is regarded as an early landmark for disease progression in prodromal HD individuals. Although a rapid cognitive decline at the early stage, attributed to the onset of HD, has been observed ~\cite{chen2022case}, an overarching study of how cognitive impairment interacts with HD onset was lacking, but will be helpful to understand the HD progression in prodromal HD individuals. In this work, we propose a holistic model to uncover the relationship between cognitive decline in multiple domains and HD onset using joint modeling of multivariate longitudinal and event time data.

Joint modeling of longitudinal and event time data has become increasingly important in biomedical studies, especially in chronic diseases, HIV/AIDS, cardiovascular research, etc, when understanding the interplay between longitudinal biomarkers and clinical events is crucial. The joint modeling framework was developed by simultaneously modeling the longitudinal and event time data under a structure with shared subject-specific random effects ~\cite{WulfsohnTsiatis1997, Henderson-etal-2000, song-etal-2002, Lin-etal-2002, Brown-etal-2005, Hsieh-etal-2006,  Li-etal-2009}. Recent development of the joint modeling method allows researchers to study the causal effect of the exposure on the survival outcome through the longitudinal marker~\cite{zhou2023causal, le2025validation}. Previous work showed that by including both the random effect and the longitudinal marker in the survival model, the joint model framework allows the researcher to separate the direct causal effect of the longitudinal marker on the outcome and the time-independent unmeasured confounding between the marker and the outcome~\cite{zheng2022quantifying, liu2018exploring}. The maximum likelihood method equipped with the EM algorithm has been a popular approach to estimate parameters in a joint model \cite{rizopoulos2012joint} for making statistical inferences. It is worth noting that most of the joint models were developed for longitudinal biomarkers and right-censored event time, emphasizing either unbiased inference for event time outcome using time-dependent longitudinal biomarkers ~\cite{song-etal-2002, Lin-etal-2002, DupuyMesbah-2002} or unbiased inference of longitudinal trajectories of the biomarkers subject to informative drop-out ~\cite{Diggle-etal-2008,  Han-etal-2014, Park-etal-2022}. These methods are not applicable to study the relationship between cognitive impairments and the HD onset in prodromal HD individuals, which is subject to interval censoring bracketed by two adjacent motor diagnostic times, where the first time shows negative and the second time yields the motor diagnosis of HD. Although joint models of longitudinal biomarkers and interval-censored event times have been studied more recently ~\cite{Gueorgruieva-etal-2012, Rouanet-etal-2016, chen2018semiparametric, Wu-Li-2021}, the methods did not concern the changes of longitudinal biomarkers triggered by the event time. Some two-phase changing-point analyses of longitudinal data around an interval-censored event time have been explored ~\cite{zhang2016robust, chu-etal-2019, chu2020stochastic}, but they did not address how the longitudinal data impacted the interval-censored event time. In addition, none of the aforementioned approaches were concerned with causal inferences.

\begin{figure}[h!]
\centering
    \includegraphics[width=0.75\columnwidth]{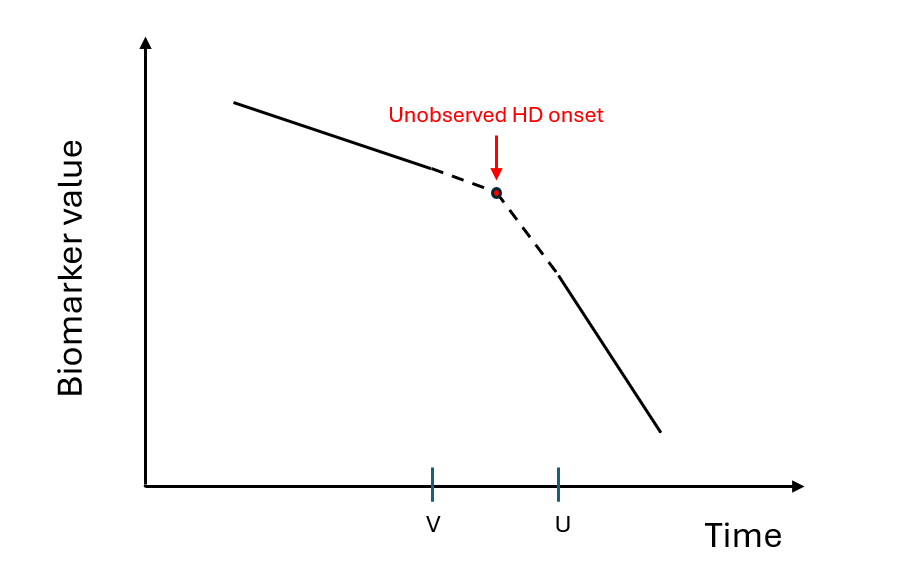}
\caption{A hypothetical model for the HD disease progression: V and U are the two adjacent diagnosed times, with V being negative and U being positive for the HD diagnosis.}
\label{fig1}
\end{figure}

In this work, we propose a two-phase approach by extending the likelihood method for joint modeling of longitudinal and interval-censored event time data to incorporate a potential change point in longitudinal biomarkers anchored at an unobserved event time, depicted in Figure (\ref{fig1}), using the causal framework for the joint modeling ~\cite{liu2018exploring}. The first phase of the model emphasizes how the longitudinal data impacts the event time, and the second phase investigates whether/how the event time changes the trajectories of longitudinal biomarkers.  We use an adaptive Newton-Raphson algorithm to compute the maximum likelihood estimates (MLE) of all coefficients in this joint model, and the nonparametric bootstrap method to estimate the standard error of all estimated coefficients.  

The rest of this paper is organized as follows: in Section \ref{sec2}, we describe our notation, models, and the likelihood method for the causal two-phase joint modeling of changing-point longitudinal data and interval-censored event time and illustrate how to calculate the MLE. In Section \ref{sec3}, we present simulation studies to evaluate the finite-sample performance of the proposed method. In Section \ref{sec4}, we apply our method to the data from the Neurobiological Predictors of Huntington's Disease Study (PREDICT-HD), a 12-year multisite prospective cohort study conducted between September 2002 and April 2014~\cite{paulsen2008detection} to examine the interplay between cognitive decline and the HD onset in prodromal HD individuals that cognitive declines predict the HD onset and the later accelerates cognitive declines. In Section \ref{sec5}, we summarize our findings and discuss potential extensions. 

\section{Methods}\label{sec2}
\subsection{Notation}
First, we define notation for the observed data. We consider $K$-dimensional longitudinal biomarker processes $M(t)=(M^{1}(t),M^{2}(t),\cdots,M^{K}(t))$ that are potentially predictive of the onset time of disease $E$, which is interval censored by $(V,U]$ via periodic diagnoses, where $V$ and $U$ can be zero and infinity, respectively. Let $X(t)$ be covariate processes that are potentially related to both longitudinal disease biomarkers and the onset time of disease. Assume the processes are assessed longitudinally at $m$ random observation times $\underline{T}=(t_1, t_2,\cdots,t_m)$, and the data from the processes collected at $\underline{T}$ are denoted as $\underline{M}=(M(t_1),M(t_2),\cdots,M(t_m))$ and $\underline{X}=(X(t_1),X(t_2),\cdots,X(t_m))$. The observed data for an individual consist of $D=(m,\underline{T};\underline{M},\underline{X};V,U)$, and we have a total of $n$ i.i.d. copies of $D$, denoted by $\underline{D}=(D_1,D_2,\cdots,D_n$).

\subsection{Causal two-phase joint model}
In studying disease progress, we hypothesize that (1) the longitudinal disease biomarkers $M(\cdot)$ are causally affect the onset time, $E$; and (2) the onset of the disease may alter the trajectories of the biomarkers. To accommodate this hypothesis, we propose to study the disease progression in two phases, described as follows: 
\begin{align}
    M^k_{\infty}(t) &= X(t)^{\top}\beta^k+Z(t)^{\top}a^k+\epsilon^k(t),\\
   \lambda(t)&=\lambda_0(t)\exp[\theta_x^{\top}X(t)+\theta_a^{\top}a^k+\theta_M^{\top}M_{\infty}(t)],\\
    M^k_{E}(t) &= M^k_{\infty}(t) + \gamma^k I(t > E)(t - E), 
\end{align}
for $k=1,2,\cdots,K$, where $M^k_{\infty}(\cdot)$ denote the $k$th potential longitudinal biomarker process shall the disease never happened and $M^k_{E}(\cdot)$ denote the $k$th potential longitudinal biomarker process shall the disease happened at time $E$. By consistent assumption, the observed $k$th longitudinal biomarker process will satisfy $M^k(t)=M_E^k(t)$ after the disease has started at time $E$ and we can divide it into two phases in the likelihood calculation, where $M^k_1(t)=M_{\infty}(t), t\in (0,E)$ and $M^k_2(t)=M^k_E(t), t\in [E, \infty)$ denote the first (before $E$) and second (after $E$) phases of the $k$th longitudinal biomarker process, $M^{k}(\cdot)$, respectively.

In the two-phase longitudinal model, $\beta^k$ represents the effects 
of the covariates on the $k$th biomarker; $a=(a^1, a^2, \cdots a^K)\sim N(0,\Sigma_a)$ represent the random effects at the individual level, which are used to surrogate time-independent unmeasured confounding between the longitudinal biomarkers and the survival outcome; $Z(\cdot)$ are the covariates of longitudinal processes corresponding to the random effects, which are often a part of $X(\cdot)$ in practice of the joint modeling; $\epsilon(t)=(\epsilon^1(t), \epsilon^2(t),\cdots, \epsilon^K(t)) \sim N(0,\Sigma_e)$ are i.i.d. time-independent random variations for observed longitudinal data; $\gamma^k$ represents the changing-point effect anchored at the interval-censored disease onset time, $E$, on the $k$th longitudinal biomarker.

For the survival model, $\lambda_0(t)$ represents an unspecified baseline hazard function for the onset time; 
$\theta_x$ represents the effects 
of the covariates on the onset time; $\theta_a$ represents the effects of the random effects in the survival model, which reflect whether unmeasured baseline variables confound the association between longitudinal biomarkers and survival outcome; $\theta_M$ represents the effects of longitudinal biomarkers at the first phase on the onset time.

The proposed two-phase joint model is a natural extension of the causal joint models for longitudinal and survival data \cite{liu2018exploring,zheng2022quantifying}, which, \textbf{however,} did not look at how disease onset impacts the trajectories of longitudinal biomarkers and only dealt with right-censored disease onset time. With an interval-censored disease onset time and two-phase longitudinal models, the likelihood of the observed data is much more complicated, with more involved numerical challenges in computation.

\subsection{The likelihood for the observed data}

In this subsection, we construct the likelihood function of the parameters $\Theta=$\\$(\lambda_0(\cdot),\beta,\gamma,\theta_X,\theta_a,\theta_M,\Sigma_a,\Sigma_e)$ of the two-phase joint models (1)-(3) for the observed data $\underline{D}$, where $\beta=(\beta^1,\beta^2,\cdots,\beta^K)$ and $\gamma=(\gamma^1,\gamma^2,\cdots,\gamma^K)$. As a conventional approach in joint modeling with shared random effects, we assume that given the shared random variables $a$, the observed data on the longitudinal biomarkers are independent within the same biomarker process and between different biomarker processes, as well as independent of the disease onset time.

First, we construct the likelihood for the data, in which the disease onset time is exactly known. We separate the observed biomarker profile $\underline{M}$ to two phases, $\underline{M}_1=$\\$(M(t_1),M(t_2),\cdots,M(V))$ (before $E$) and $\underline{M}_2=(M(U),\cdots,M(t_m))$ (after $E$). The joint model with shared random effects (1)-(3) gives
\begin{eqnarray}
f(E,\underline{M}|m,\underline{T},\underline{X},a)= f(\underline{M}_2|E,m,\underline{T},\underline{X},a)f(E|\underline{M}_1,m,\underline{T},\underline{X},a)f(\underline{M}_1|m,\underline{T},\underline{X},a),
\end{eqnarray}
where 
\begin{align*}   f(E|\underline{M}_1,m,\underline{T},\underline{X},a)=&\lambda_0(E)\exp(\theta_X^{\top}X(E)+\theta_a^{\top}a+\theta_M^{\top}M_1(E))\\
    &\times \exp(\big\{-\sum_{j:t_j\le E}\int_{t_{j-1}}^{t_j}\lambda_0(s)\exp(\theta_X^{\top}X(t_{j-1})+\theta_a^{\top}a+\theta_M^{\top}M_1(t_{j-1}))ds\big\}),
\end{align*}
for which we extrapolate the values of the processes $X(\cdot)$ and $M_1(\cdot)$ between the two observation times $t_{j-1}$ and $t_j$ by their values at $t_{j-1}$ (we denote $t_0$ as 0), respectively, because the processes are only observed at the discrete times, $\underline{T}$,
\begin{align*}
f(\underline{M}_1|m,\underline{T},\underline{X},a) 
&= \prod_{j:t_j<E} \Biggl[ \frac{1}{\sqrt{(2\pi)^K\det(\Sigma_e)}} \times \exp\Biggl(-\frac{1}{2}\Big(M_1(t_j)-X(t_j)^{\top}\beta-Z(t_j)^{\top}a\Big)^{\top}\Sigma_e^{-1} \\
&  \times \Big(M_1(t_j)-X(t_j)^{\top}\beta-Z(t_j)^{\top}a\Big)\Biggr) \Biggr]
\end{align*}
and
\begin{align*}
f(\underline{M}_2|E,m,\underline{T},\underline{X},a) 
&= \prod_{j:t_j\geq E} \Biggl[ \frac{1}{\sqrt{(2\pi)^K\det(\Sigma_e)}} \\
& \quad \times \exp\Biggl(-\frac{1}{2}\Big(M_2(t_j)-X(t_j)^{\top}\beta-Z(t_j)^{\top}a-\gamma (t_j-E)\Big)^{\top}\Sigma_e^{-1} \\
& \qquad \qquad \times \Bigl(M_2(t_j)-X(t_j)^{\top}\beta-Z(t_j)^{\top}a-\gamma (t_j-E)\Bigr)\Biggr)\Biggr].
\end{align*}

Since $E$ is bracketed by the interval $(V,U]$ in our study of the disease progression for prodromal HD subjects, where $U$ is possibly infinity for subjects who had not been diagnosed at the end of the follow-up, we introduce a binary indicator, $\Delta$, such that $\Delta=1$ for the case that $E$ is within $(V,U]$ made by two subsequent observation times and $\Delta=0$ for $E$ being right censored at the end of follow-up, i.e. $V=t_m$. Hence, 
\begin{eqnarray*}
&&f(\underline{M},V<E\le U|m,\underline{T},\underline{X},a)=\left[\int_V^U f(\underline{M}_2|t,m,\underline{T},\underline{X},a)f(t|\underline{M}_1,m,\underline{T},\underline{X},a)dt\right]^{\Delta} \\
&& \hspace{15mm} \times S(V|\underline{M}_1,m,\underline{T},\underline{X},a)^{1-\Delta}f(\underline{M}_1|m,\underline{T},\underline{X},a),
\end{eqnarray*}
where
\[
S(V|\underline{M}_1,m,\underline{T},X,a)
    = \exp\left\{ -\sum_{j:t_j\leq V}\int_{t_{j-1}}^{t_j}\lambda_0(s)\exp(\theta_x^{\top}X(t_{j-1})+\theta_a^{\top}a+\theta_m^{\top}M_1(t_{j-1}))ds\right\}.
\]
Assuming that the number of observations, observation times, and the covariate processes are noninformative to the model parameters $\Theta$, the likelihood for $D$ as a function of $\Theta$ is
\begin{eqnarray*}
L(\Theta;D)=&&\int \left[\int_V^U f(\underline{M}_2|t,m,\underline{T},\underline{X},a)f(t|\underline{M}_1,m,\underline{T},\underline{X},a)dt\right]^{\Delta} \\&& \times S(V|\underline{M}_1,m,\underline{T},\underline{X},a)^{1-\Delta}f(\underline{M}_1|m,\underline{T},\underline{X},a)f(a;\Sigma_a)da,
\end{eqnarray*}
where
\[
f(a;\Sigma_a) = (2\pi)^{-K/2}(\det(\Sigma_a))^{-1/2}\exp(-\frac{1}{2}a^{\top}\Sigma_a^{-1}a),
\]
and the likelihood for the observed data $\underline{D}$ is
\begin{eqnarray}
&& L(\Theta;\underline{D})=\prod_{i=1}^{n}L(\Theta;D_i)= \prod_{i=1}^{n}\int \left[\int_{V_{i}}^{U_{i}} f(\underline{M}_{i2}|t,m_i,\underline{T}_i,\underline{X}_i,a_i)f(t|\underline{M}_{i1},m_i,\underline{T}_i,\underline{X}_i,a_i)dt\right]^{\Delta_i} \nonumber  \\
&& \hspace{15mm}\times S(V_i|\underline{M}_{i1},m_i,\underline{T}_i,\underline{X}_i,a_i)^{1-\Delta_i}f(\underline{M}_{i1}|m_i,\underline{T}_i,\underline{X}_i,a_i)f(a_i;\Sigma_a)da_i.
\end{eqnarray}

We implemented a spline-based seminparametric sieve maximum likelihood method to estimate the model parameters $\Theta$ by approximating the baseline cumulative hazard function $\Lambda_0(t)=\int_0^t\lambda_0(u)du$ using the cubic monotone B-spline 
\[
\Lambda_0(t)=\sum_{j=1}^{q_n}\alpha_jB_j(t), \ \ 0\le \alpha_1\le \alpha_2\le \cdots \le \alpha_{q_n}.
\]
where $\{\alpha_j\}_{j=1}^{q_n}$ are the coefficients of B-spline that are non-negative and monotone non-decreasing, and $q_n$ is the number of basis functions that depends on sample size. The monotonicity of the B-spline coefficients warrants the monotone non-decreasing property of $\Lambda_0(\cdot)$ \cite{schumaker2007spline}. The spline-based semiparametric sieve maximum likelihood estimation method has been widely employed for many semiparametric models ~\cite{lu2009semiparametric,zhang2010spline,hua2014spline,su2024semiparametric}.  

\subsection{Numerical Algorithm}
For the spline-based semiparametric sieve maximum likelihood estimation, the model parameters become
\[
\Theta=(\alpha,\beta,\gamma,\theta_X,\theta_a,\theta_M,\Sigma_a,\Sigma_e)
\]
with $\alpha = (\alpha_1,\alpha_2,\cdots,\alpha_{q_n})$ being non-negative and monotone non-decreasing. We adopted a general Fisher scoring algorithm with a line search procedure to compute the estimates:
\[
\Theta^{(p+1)}=\Theta^{(p)}-\eta I^{-1}(\Theta^{(p)})\dot{l}(\Theta^{(p)}),
\]
where $\dot{l}(\Theta^{(p)})=\nabla_{\Theta}l(\Theta^{(p)};\underline{D})=\sum_{i=1}^n\nabla_{\Theta}l(\Theta^{(p)};D_i)=\sum_{i=1}^n\nabla_{\Theta} \log L(\Theta^{(p)};\underline{D}_i)$ represents the score of the log likelihood function, and $I(\Theta^{(p)})=\sum_{i=1}^n\dot{l}(\Theta^{(p)};D_i)\dot{l}(\Theta^{(p)};D_i)^T$ represents the observed information matrix, where $\dot{l}(\Theta^{(p)};D_i)=\nabla_{\Theta}l(\Theta^{(p)};D_i)=\nabla_{\Theta}\log L(\Theta^{(p)}; D_i)$. $\eta$ is an adaptive step length when updating the parameters. We used the step-halving adaptive line search strategy to ensure that the log likelihood increases after each iteration. We stopped the algorithm when the maximum of relative differences of parameters was less than $10^{-3}$, i.e. $\max_{\Theta} |\Theta^{(p+1)}-\Theta^{(p)}|/|\Theta^{(p)}|<10^{-3}$.

While the algorithm appears to be straightforward, several challenges in numerical implementation were noted and were practically handled. First, it was hard to perform Fisher scoring algorithm with monotone constraints on $\{\alpha_j\}_{j=1}^{q_n}$. We used a reparameterization approach to remove the constraints in optimization by letting $\alpha_j = \sum_{l=1}^j\exp(\xi_l)$, for $j=1,2,\cdots,q_n$. 
Second, there was no closed form for the integration over the interval $(V_i, U_i)$ since the function to be integrated includes a product of exponential of cubic B-splines and a polynomial function of time $t$. 
To calculate this integral as accurately as possible, we choose to use Gauss-Legendre quadrature, which has been proven to be reliable for approximating the bounded integral of a smooth function. We used 20 nodes inside $(V_i, U_i)$ for the Gauss-Legendre quadrature computation.
There is also no explicit form of the integration over random effects $a$. We used the Gauss-Hermite quadrature, which is particularly designed for approximating the value of integrals of functions with form $e^{-x^2}f(x)$. We also used 20 nodes for the Gauss-Hermite quadrature.
Third, it is a daunting job to get the score function and Hessian matrix for the likelihood (5). We used the finite difference method to approximate the score:
\[
\dot{l}^*_{u}(\Theta^{(p)}) = \frac{l(\Theta^{(p)}+\delta\cdot 1_u;\underline{D}) - l(\Theta^{(p)}-\delta\cdot 1_u;\underline{D})}{2\delta},
\]  
where $1_u$ is a vector with 1 for the $u$-th element and 0 for the other elements, and $\dot{l}^*_u$ is the finite difference of the log likelihood with $\delta$ chosen to be a very small value in a magnitude of $10^{-6}$ in our calculation. Consequently, the observed Fisher information matrix was approximated by $I^*(\Theta^{(p)})=\sum_{i=1}^n\dot{l}^*(\Theta^{(p)};D_i)\dot{l}^*(\Theta^{(p)};D_i)^T$. 
Last, the eigenvalues of observed information matrix $I^*(\Theta^{(p)})$ could be very small during the numerical iteration, causing trouble to invert the approximated Fisher information matrix $I^*(\Theta^{(p)})$. We used the generalized inverse matrix of $(I^*)^-(\Theta^{(p)})$ instead, for the Fisher scoring algorithm.

For the initial values of $\Theta$, we fitted longitudinal and survival models separately. First, we fitted a longitudinal model for each biomarker with fixed and random effects and the time effect after event by assuming $E=(U+V)/2$. 
Then, we fitted a Cox model by assuming $\frac{U+V}{2}$ as a right-censored event time, and included both fixed covariates and estimated random effects from the longitudinal model as predictors. We used the JM packages ~\cite{rizopoulos2012joint} to combine the longitudinal and survival models to get an initial joint estimate of $\Theta$ except the baseline hazard, using the estimated parameters obtained above. To give an initial value of B-spline cumulative hazard function, we first pre-specify the number of knots $q_n$ in the order of $q^{1/3}$ with $q$ being the total number of the distinct values in the set of $\{U_i,V_i|i=1,\dots,n\}$  ~\cite{zhang2016robust}, we used the corresponding quantiles, $t_1,t_2,\dots,t_{q_n-4}$ of the distinct values in the set of $\{U_i,V_i|i=1,\dots,n\}$, as the interior knots and applied the Maximum Likelihood Estimation (MLE) method for interval-censored survival data and \textit{optim} function in R to complete the initial values of $\xi$.

To estimate the standard errors of the estimated $\Theta$, we used the nonparametric Bootstrap method with $B=50$ repetitions. 

\section{Simulation}\label{sec3}
\subsection{Settings}
We conducted a comprehensive simulation study to evaluate the performance of our proposed two-phase model. For the $i$th subject, the observed data were generated from the following specific model of the two-phase joint model (1)-(3):
\begin{align*}
            M_{\infty,i}^k(t) &= \beta_{0}^k+\beta_{1}^k t+\beta_{2}^k X_i +a_{i}^k+\epsilon_i^k(t),\\            
            \lambda_i(t) &= \lambda_0(t)\exp(\theta_x^{\top}X_i+\theta_a^{\top}a_i+\theta_m^{\top}M_{\infty,i}(t)), \\
            M_{E,i}^k(t) &= M_{\infty,i}^k(t) + \gamma^k I(t > E_i)(t - E_i), 
\end{align*}
where we considered 2 biomarkers ($k$=1,2) for the longitudinal model. A single time independent covariate was simulated according to $X_i\sim N(0,0.8)$,  
and two random effects $a_i=(a^1_i,a^2_i)$ were generated from a joint normal distribution $a_i\sim N(0, \Sigma_a)$.

For the simulation study, we set $\beta_0=(5,5)^{\top}, \beta_1=(-0.2,-0.2)^{\top}, \beta_2=(-0.3,-0.3)^{\top}, \gamma=(-0.4,-0.4)^{\top}, \theta_x=0.3, \theta_a=(0.1,0.1)^{\top},\theta_m=(-0.2,-0.2)^{\top},\Sigma_a =\begin{pmatrix}0.5 &0.1\\0.1 &0.5
\end{pmatrix}, \Sigma_e=\begin{pmatrix}0.5 &\frac{\sqrt{2}}{10}\\ \frac{\sqrt{2}}{10} &1
\end{pmatrix}$, and $\lambda_0(t)=0.3\sqrt{t/5}$ to generate the data.
$m=11$ observations were simulated from the aforementioned setting for each individual, 
with the observation times $\underline{T}_i=\{t_{ij}\}_{j=1}^{10}$ independently generated from uniform distribution $U(j-0.2, j+0.2), j=1,2,\dots,10$. 

In the simulation study, we used the cubic B-splines with 6 and 10 interior knots to carry out the spline-based semiparametric maximum sieve likelihood method to estimate the model parameters $\Theta$, respectively, to evaluate how the analysis results are sensitive to the selection of the number of spline knots. 
We ran 1,000 repeated trials with sample sizes of 400 and 800, respectively, to evaluate the finite sample performance based on the asymptotic normality theory. Estimated bias, SD (Monte-Carlo sample standard deviation), ASE (Average of the Bootstrap standard error estimates), and CP (Coverage Probability of the 95\% Wald confidence interval) are reported in the results. 
The simulation study was conducted using the Holland Computer Center (HCC) server at the University of Nebraska, Lincoln. All codes were written in the R module 4.4. The simulation and application code can be found in the \href{https://github.com/Emomeow/PhD-Thesis}{\textit{Github respiratory}}. The predict-HD data can be found at \href{https://www.ncbi.nlm.nih.gov/projects/gap/cgi-bin/dataset.cgi?study_id=phs000222.v3.p2&phv=88237&phd=2387&pha=&pht=3448&phvf=&phdf=&phaf=&phtf=&dssp=1&consent=&temp=1}{\textit{NIH}}.

\subsection{Results}

Table \ref{table2} displays the results of the spline-based semiparametric maximum sieve likelihood analysis for regression coefficients, using the B-spline approximation with 6 interior knots for the baseline cumulative hazard function. It appears that the estimation biases of the regression coefficients were negligible, the averages of Bootstrap standard error estimates were all very close to the corresponding Monte-Carlo standard deviations, indicating that the Bootstrap estimation of standard error was reliable, and the 95\% Wald confidence intervals yielded a right coverage probability of 0.95.
When the sample size increased from 400 to 800, all the standard deviations and Bootstrap standard errors decreased by approximately $1/\sqrt{2}$. These data provided strong evidence for the validity of inference based on the asymptotic normality theory.

\begin{figure}[h!]
        \begin{minipage}[b]{0.48\textwidth}
  \centering
  \includegraphics[width=\textwidth]{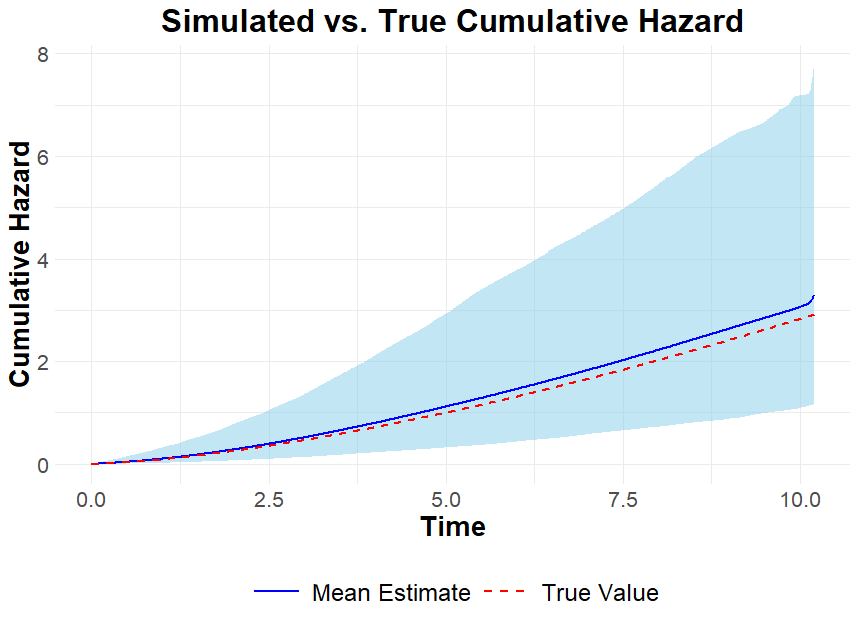}
  \caption{Estimated B-spline cumulative hazard with interior knots=6, n=400, true $\Lambda_0(t)=(0.2t)^{1.5}$.} 
  \label{fig4}
\end{minipage}
\hfill
\begin{minipage}[b]{0.48\textwidth}
  \centering
  \includegraphics[width=\textwidth]{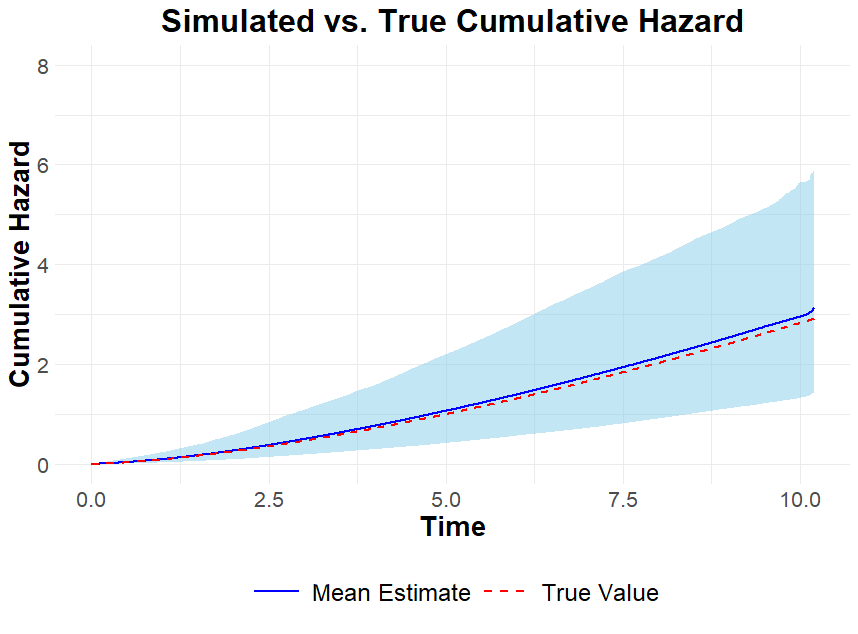}
  \caption{Estimated B-spline cumulative hazard with interior knots=6, n=800, true $\Lambda_0(t)=(0.2t)^{1.5}$.} 
  \label{fig5}
\end{minipage}
\end{figure}

Figures \ref{fig4} and \ref{fig5} plot the estimated cumulative hazard function, using the B-spline with 6 interior knots, for sample sizes of cases $n=400$ and $800$, respectively. The red dashed curve is the true cumulative hazard. The blue solid curve was the pointwise average of the estimated cumulative hazard over the 1000 trials. The light-blue band refers to the pointwise 95\% quantile region, where the lower and upper ends are the 2.5\% and 97.5\% percentiles of the pointwise estimates of the cumulative hazard function, respectively, over the 1,000 repeated trials. It is evident that the B-spline estimate of the cumulative hazard function was consistent, and the standard error of the estimation decreased as the sample size increased.

The simulation results (not shown here) with 10 interior knots in the cubic B-spline estimation for the baseline cumulative hazard function $\Lambda_0(\cdot)$ were similar to those shown in Table 1 and Figures 2 and 3, indicating that the proposed spline-based semiparametric maximum sieve likelihood analysis is insensitive to the choice of the number of spline knots when $\Lambda_0(\cdot)$ is smooth. 

\begin{table}[h!]
\centering
\begin{tabular}{clrrrr}
n=400                        & Parameters & Bias   & SD     & ASE    & CP    \\
Biomarker 1 & $\beta_0$  & -0.002  & 0.042  & 0.044  & 0.956 \\
                             & $\beta_1$  & .000  & 0.004  & 0.004  & 0.949 \\
                             & $\beta_2$  & .000 & 0.043  & 0.045  & 0.935 \\
                             & $\gamma$   & .000   & 0.011  & 0.011  & 0.947 \\
Biomarker 2 & $\beta_0$  & 0.002  & 0.045  & 0.046  & 0.949 \\
                             & $\beta_1$  & -.000  & 0.006  & 0.006  & 0.937 \\
                             & $\beta_2$  & -0.002 & 0.044  & 0.043  & 0.949 \\
                             & $\gamma$   & .000   & 0.015  & 0.015  & 0.940 \\
Survival Model & $\theta_x$    & 0.011  & 0.090 & 0.094 & 0.954 \\
                                & $\theta_{m1}$ & 0.005  & 0.112 & 0.114 & 0.950 \\
                                & $\theta_{m2}$ & 0.001  & 0.082 & 0.080 & 0.940 \\
                                & $\theta_{a1}$ & -0.005  & 0.167 & 0.175 & 0.955 \\
                                & $\theta_{a2}$ & 0.002 & 0.154 & 0.161 & 0.952 \\
n=800                        & Parameters & Bias   & SD     & ASE    & CP    \\
Biomarker 1 & $\beta_0$  & 0.001 & 0.031  & 0.031  & 0.943 \\
                             & $\beta_1$  & .000   & 0.003  & 0.003  & 0.944 \\
                             & $\beta_2$  & -0.001  & 0.040  & 0.040  & 0.941 \\
                             & $\gamma$   & .000   & 0.009  & 0.009  & 0.936 \\
Biomarker 2 & $\beta_0$  & -.000  & 0.039 & 0.040 & 0.940 \\
                             & $\beta_1$  & .000   & 0.006 & 0.006 & 0.951 \\
                             & $\beta_2$  & 0.001  & 0.029 & 0.031 & 0.953 \\
                             & $\gamma$   & .000  & 0.008 & 0.008 & 0.939 \\
Survival Model & $\theta_x$    & 0.006  & 0.069 & 0.066 & 0.924 \\
                                & $\theta_{m1}$ & 0.002  & 0.082 & 0.080 & 0.939 \\
                                & $\theta_{m2}$ & -0.001  & 0.059 & 0.057 & 0.936 \\
                                & $\theta_{a1}$ & 0.003 & 0.128 & 0.122 & 0.936 \\
                                & $\theta_{a2}$ & -0.003 & 0.111 & 0.112 & 0.951
\end{tabular}
\caption{Simulation results for joint model using B-spline cumulative hazard with 6 interior knots.}
\label{table2}
\end{table}

\section{A REAL-WORLD APPLICATION: PREDICT-HD STUDY}\label{sec4}
\subsection{Data and Study Objectives}

We applied our proposed causal two-phase joint model to data from the Neurobiological Predictors of Huntington's Disease (PREDICT-HD) study, a 12-year multisite prospective cohort study conducted from September 2002 to April 2014 \cite{paulsen2008detection}. The study recruited 1,155 individuals with HD gene expansion and 317 controls without HD gene expansion, primarily family members of HD gene–expanded individuals. The primary objective of this analysis was to use our method to empirically test two central hypotheses: (1) that cognitive decline predicts time to motor onset in prodromal HD, and (2) that motor onset acts as a change point that accelerates the rate of subsequent cognitive decline.


The PREDICT-HD data, publicly available in dbGaP PREDICT-HD Huntington's Disease Study (nindsdac@mail.nih.gov), are uniquely suited to our model because the clinical diagnosis of HD was determined by a certified motor assessor trained by the Huntington Study Group (HSG), who evaluated each participant using the Diagnostic Confidence Level (DCL) \cite{kieburtz1996unified}. A DCL score of 4 indicates greater than 99\% confidence that observed motor abnormalities are definitive signs of HD. Because diagnoses were made at discrete clinical visit times, this process naturally resulted in an interval-censored HD onset time, denoted by $(V,U]$, for individuals diagnosed during the study.

We selected two key longitudinal biomarkers in cognitive domains \cite{harrington2012cognitive}: the Symbol Digit Raw Score Total (\textit{sydigtot}), which measures processing speed and attention, and the Stroop Word Reading Total (\textit{stroopwo}), which assesses information integration and inhibition. We included four time-independent baseline covariates: age at study entry, education (in years), gender (0=Female, 1=Male), and the baseline HD disease burden (CAP) score. The CAP score, which combines age and CAG repeat length, is a well-established prognostic measure~\cite{zhang2021mild}. For scaling purposes, the CAP score was divided by 100.

After removing individuals diagnosed at their first visit or with only a single observation, the final analysis dataset comprised 983 prodromal HD individuals with a total of 5,694 longitudinal observations. 


\subsection{Results}
We fit the proposed causal two-phase joint model using a cubic B-spline with 6 interior knots for the baseline cumulative hazard function. The results, presented in Table \ref{table3} (Longitudinal Model) and Table \ref{table4} (Survival Model), provide strong quantitative evidence supporting our hypotheses.

The survival component of the joint model (Table~\ref{table4}) evaluates the association between early cognitive decline and time to the motor HD onset. The results support our first hypothesis. After adjusting for CAP, \textit{sydigtot} was a significant predictor of HD onset, with a hazard ratio (HR) of 0.955 (95\% CI: (0.918, 0.993), $p = 0.021$). In contrast, \textit{stroopwo} was a weaker predictor, with an HR of 0.981 (95\% CI: (0.962, 1.001), $p = 0.0624$). Hazard ratios less than 1.0 indicate that higher cognitive scores (i.e., better performance) are associated with a lower risk of motor onset at any given time, whereas poorer cognitive performance is associated with an increased hazard and earlier HD onset.

As expected, baseline CAP score was the primary driver of HD onset (HR = 2.159, 95\% CI: (1.578, 2.954), $p < 0.0001$), confirming that greater baseline disease burden strongly predicted HD onset. We also observed a significant effect of sex (HR = 0.737, $p = 0.0493$), suggesting that males had a lower hazard of HD onset in this cohort.

The longitudinal component of the joint model (Table~\ref{table3}) characterizes the two-phase progression of cognitive biomarkers, with the unobserved motor onset $E$ serving as the change point. The results provide strong support for our second hypothesis. The change-point effect ($\gamma$) was highly significant and negative for both \textit{sydigtot} ($\hat{\gamma} = -1.288$, 95\% CI: (-1.657, -0.919), $p < .0001$) and \textit{stroopwo}($\hat{\gamma} = -2.206$, 95\% CI: (-2.817, -1.596), $p < .0001$). These findings offer clear, data-driven validation of the two-phase progression illustrated in Figure~\ref{fig1}.

For \textit{stroopwo}, for example, cognitive performance declined at a baseline rate of 0.848 units per year prior to motor diagnosis. Following HD onset, the rate of decline accelerated by an additional 2.206 units per year, yielding a total post-onset decline of 3.054 units per year. This demonstrates that HD onset is not merely a concurrent clinical milestone, but a critical event marking a substantial acceleration in the deterioration of cognitive function.

The covariate effects in the longitudinal model were also clinically consistent. Higher baseline CAP scores (greater disease burden) and older age were associated with significantly lower cognitive scores. In contrast, higher levels of education were associated with significantly higher cognitive scores, suggesting a protective effect.

A key finding from the survival model (Table~\ref{table4}) was the non-significance of the random effects terms (random \textit{sydigtot}, $p = 0.976$; random \textit{stroopwo}, $p = 0.592$). In the joint modeling framework \cite{zheng2022quantifying}, these terms capture unmeasured time-independent confounding. Their non-significance suggests that, after accounting for the observed cognitive trajectory, $M_1(t)$, the underlying shared variability (i.e., the random intercepts $a^k$) does not provide additional predictive power for HD onset. This reinforces the interpretation that the association is driven by the cognitive decline process itself, rather than by an unmeasured confounder influencing both cognition and motor onset.

In summary, applying our model to the PREDICT-HD cohort successfully disentangled the complex, bidirectional relationship between cognitive and motor symptoms. The results provide strong statistical evidence that cognitive decline is a significant harbinger of HD onset, and that HD onset, in turn, serves as a critical change point that accelerates subsequent cognitive deterioration.

\begin{table}[h!]
\centering
\begin{tabular}{clrrrr}
\multirow{2}{*}{}                                                         & \multirow{2}{*}{Coefficients} & \multirow{2}{*}{Estimate} & \multicolumn{2}{c}{Bootstrap 95\% CI*} & \multirow{2}{*}{p-value} \\
                                                                          &                               &                           & \multicolumn{1}{c}{Lower}   & Upper    &                          \\
\begin{tabular}[c]{@{}c@{}}Longitudinal model\\ for sydigtot\end{tabular} & Intercept                     & 57.773                    & 50.347                      & 65.198   & \textless{}.0001         \\
                                                                          & Time                          & -0.550                    & -0.632                      & -0.467   & \textless{}.0001         \\
                                                                          & Age                           & -0.139                    & -0.227                      & -0.051   & 0.0019                   \\
                                                                          & Education                     & 1.349                     & 0.973                       & 1.725    & \textless{}.0001         \\
                                                                          & Gender                        & -1.809                    & -3.979                      & 0.361    & 0.1022                   \\
                                                                          & CAP/100                       & -5.559                    & -6.393                      & -4.725   & \textless{}.0001         \\
                                                                          & Time after diagnosis          & -1.288                    & -1.657                      & -0.919   & \textless{}.0001         \\
\begin{tabular}[c]{@{}c@{}}Longitudinal model\\ for stroopwo\end{tabular} & Intercept                     & 102.683                   & 92.847                      & 112.519  & \textless{}.0001         \\
                                                                          & Time                          & -0.848                    & -0.984                      & -0.712   & \textless{}.0001         \\
                                                                          & Age                           & -0.061                    & -0.180                      & 0.059    & 0.3182                   \\
                                                                          & Education                     & 1.405                     & 0.797                       & 2.013    & \textless{}.0001         \\
                                                                          & Gender                        & -0.362                    & -3.253                      & 2.529    & 0.8061                    \\
                                                                          & CAP/100                       & -6.198                    & -7.625                      & -4.771   & \textless{}.0001         \\
                                                                          & Time after diagnosis          & -2.206
                    & -2.817                      & -1.596   & \textless{}.0001         \\
*CI: Confidence Interval                                                  &                               &                           & \multicolumn{1}{c}{}        &          &                         
\end{tabular}
\caption{Analysis results of longitudinal model using B-spline cumulative hazard with 6 interior knots for PREDICT-HD data.} 
\label{table3}
\end{table}

\begin{table}[H]
\centering
\begin{tabular}{clrrrr}
                   &                                &                       & \multicolumn{2}{c}{Bootstrap 95\% CI} &                                                  \\
\multirow{-2}{*}{} & \multirow{-2}{*}{Coefficients} & \multirow{-2}{*}{HR*} & \multicolumn{1}{c}{Lower}   & Upper   & \multirow{-2}{*}{p-value}                        \\
Survival model     & Age                            & 0.998                 & 0.982                       & 1.013   & 0.7646                                           \\
                   & Education                      & 1.013                 & 0.941                       & 1.092   & 0.7293                                           \\
                   & Gender                         & 0.737                 & 0.543                       & 0.999   & 0.0493 \\
                   & CAP/100                        & 2.159                 & 1.578                       & 2.954   & \textless{}.0001                                 \\
                   & sydigtot                       & 0.955                 & 0.918                       & 0.993   & 0.0210                                           \\
                   & stroopwo                       & 0.981                 & 0.962                       & 1.001   & 0.0624                                           \\
                   & random sydigtot                & 1.001                 & 0.951
                                    & 1.053   & 0.9764
                           \\
                   & random stroopwo                & 1.007                 & 0.983
                                    & 1.031   & 0.5915
                           \\
*HR: Hazard Ratio  &                                &                       & \multicolumn{1}{c}{}        &         &                                                 
\end{tabular}
\caption{Analysis results of survival model using B-spline cumulative hazard with 6 interior knots for PREDICT-HD data.} 
\label{table4}
\end{table}

\section{Final Remarks}\label{sec5}
Using the spline-based sieve likelihood approach, the proposed causal two-phase joint model produced reliable inference for the regression parameters, as demonstrated by the simulation study. The maximum likelihood estimates (MLEs) were consistent for the regression parameters, the 95\% confidence intervals achieved appropriate coverage probabilities, the standard errors decreased with increasing sample size at the expected $\sqrt{n}$ rate, and the bootstrap method provided consistent estimates of the standard errors.

Applying our proposed method to the PREDICT-HD data, we successfully characterized disease progression in prodromal HD individuals. Our analysis showed that both cognitive functions—processing speed/attention (\textit{sydigtot}) and information integration/inhibition (\textit{stroopwo})—deteriorate over time and \textit{sydigtot} is predictive of HD onset. Moreover, the rate of decline in both cognitive domains accelerated after motor onset. These findings provide novel and informative insights into the progression of HD, offering valuable guidance for the research community in characterizing disease dynamics.

While the proposed method modeled only a linear time effect to facilitate interpretation of the change-point effect in the HD study, it can be readily extended to accommodate a nonlinear time effect, specified as
\begin{equation*}
    M_E(t) = M_{\infty}(t) +I(t>E)g(t-E),
\end{equation*}
where $g()$ is a predetermined smooth function of time after the disease onset. Since the function to be integrated between interval $(V,U): f(M_2|t,X,Z,a;\beta,\gamma,\Sigma_e)f(t|M_1,X,Z,a;\theta,\lambda))$ is still smooth, the Gauss-Legendre quadrature remains to be an effective method to approximate the integration in computing the likelihood.



Our proposed method did not account for measurement error in the longitudinal biomarkers. As a result, the residual term reflected the total biomarker variation over time contributing to the risk of disease in the survival component of the joint model. In practice, however, observed longitudinal biomarkers are often subject to measurement error, meaning that only a portion of the residual represents the true underlying variation. To address this, the measurement error component should be excluded in the hazard model by replacing the observed marker with an error-free version,$M^{k*}_{\infty}(t)$, to replace the observed marker $M^{k}_{\infty}(t)$.
This will generalize our model to the following:
\begin{align*}
    M^{k*}_{\infty}(t) &= X(t)^{\top}\beta^k+Z(t)^{\top}a^k+\epsilon^k(t), \\
    M^k_{\infty}(t) &= M^{k*}_{\infty}(t) + e^k(t),\\
    M^k_E(t) &= M^k_{\infty}(t) + \gamma^k I(t > E)g(t - E), \\
    \lambda(t)&=\lambda_0(t)\exp[\theta_x^{\top}X(t)+\theta_a^{\top}a^k+\theta_M^{\top}M^*_{\infty}(t)],
\end{align*}
for $k =1,2, \dots,K$, where $M^{k*}_{\infty}(t)$ denotes the true value of $k$-th biomarker at time $t$ in the first phase, which is unobserved in data. $M^{k}_{\infty}(t), M^k_E(t)$ denote the error-prone $k$-th biomarker's value at time $t$. $e^k(t)$ denotes independent measurement error of $k$-th biomarker at each time $t$, usually from a normal distribution. And the biomarker term in the survival model will be the true but unobserved biomarkers' values. Repeated measures or external data will be needed to estimate the variance of the measurement error in order to separate true biomarker variation and pure measurement error.

\section*{Acknowledgements}
The project described is supported by the National Institute of General Medical Sciences, U54 GM115458, which funds the Great Plains IDeA-CTR Network and the authors of this manuscript. The content is solely the responsibility of the authors and does not necessarily represent the official views of the NIH. 

\section*{Conflict of interest}

The authors declare no conflict of interest. 


\selectlanguage{english}
\FloatBarrier
\bibliography{bibliography/converted_to_latex.bib%
}

\end{CJK}

\end{document}